\documentstyle[12pt,epsfig]{article}

\large
\newcommand{\be}{\begin{eqnarray}}
\newcommand{\ee}{\end{eqnarray}}
\newcommand\del{\partial}
\newcommand{\la}{\lambda}
\newcommand{\Dirac}{\rlap {\hspace{-0.5mm} \slash} D}

\begin{document}
\setlength{\baselineskip}{17pt}
\pagestyle{empty}
\vfill
\eject
\begin{flushright}
SUNY-NTG-98/62\\
NBI-HE-98-38
\end{flushright}
\vskip 2.0cm
\centerline{\Large \bf The Microscopic Spectral Density}
\vskip 0.4cm
\centerline{\Large \bf of the QCD Dirac Operator} 

\vskip 1.2cm
\centerline{P.H. Damgaard$^1$, J.C. Osborn$^2$, D. Toublan$^2$, 
J.J.M. Verbaarschot$^2$ }
\vskip 0.2cm
\centerline{\it $^1$Niels Bohr Institute, Blegdamsvej 17, Copenhagen, Denmark}
\centerline{\it $^2$Department of Physics and Astronomy, SUNY, 
Stony Brook, New York 11794}
\vskip 1.5cm

\centerline{\bf Abstract}
We derive the microscopic spectral density of the Dirac operator
in $SU(N_c\geq 3)$ Yang-Mills theory coupled to $N_f$ fermions in
the fundamental representation. 
An essential technical ingredient is an exact rewriting of this density 
in terms of integrations over the super Riemannian manifold $Gl(N_f+1|1)$.
The result agrees exactly with earlier calculations based on
Random Matrix Theory.
\vskip 0.5cm
\noindent
{\it PACS:} 11.30.Rd, 12.39.Fe, 12.38.Lg, 71.30.+h 
\\  \noindent
{\it Keywords:} QCD Dirac operator; Chiral random matrix theory; Partially
quenched chiral perturbation theory; super-Riemannian manifolds; Microscopic 
spectral density; Valence quark mass dependence

\vfill
\noindent

\eject
\pagestyle{plain}

\vskip1.5cm
\noindent
{\bf 1. Introduction}
\vskip 0.5cm

For the understanding of chiral symmetry breaking (and restoration)
in QCD and other QCD-like theories with massless or near-massless quarks 
it is essential to know  the distribution of the small eigenvalues of the 
Dirac operator. 
In the conventional thermodynamic limit where the space-time 
volume $V$ is taken to infinity prior to taking quark masses to zero, 
the accumulation of eigenvalues $\la$ near $\la \sim 0$ 
may or may not lead 
to a non-vanishing density of the eigenvalues at the origin, $\rho(0)/V$.
According to the Banks-Casher relation \cite{BC},  
$\Sigma_0 \equiv \langle\bar{\psi}\psi\rangle
= \pi\rho(0)/V$, this spectral density is an order
parameter for chiral symmetry breaking. In all that follows we only consider
the case where the chiral condensate does not vanish in the chiral limit.

The full spectral density $\rho(\la)$ is prohibitively difficult to
compute, sensitive as it is to physics at all scales from the infrared
up to the ultraviolet cut-off. Because of its relation to the chiral
condensate, even just computing $\rho(0)$ is tantamount to 
understanding in all detail the dynamics that underlies  chiral symmetry 
breaking in QCD. Moreover, since $\rho(0)$ involves an energy scale in
the deepest infrared, only non-perturbative techniques such
as lattice regularizations can have a hope of computing this single
number from first principles. The fact that only low-momentum modes
of the Dirac spectrum are of relevance here nevertheless opens up the
interesting possibility of studying the problem in a new setting, using the
low-momentum representation of the QCD partition function. This 
effective theory is a chiral Lagrangian. Its equivalence with
the conventional representation of QCD in terms of microscopic
degrees of freedom (quarks and gluons) becomes exact in precisely the limit
of interest here: zero (or almost-zero) 
momentum\footnote{The domain of
validity of the zero momentum mode approximation is determined by the
pion decay constant $F$ \cite{Osbornprl}.}. 
There is only one input 
parameter in this approach: the chiral condensate
$\Sigma_0$, 
and hence the value of spectral density evaluated at the origin, $\rho(0)$.
For computational
purposes it is convenient to impose a condition which ensures that
only exact zero-momentum modes dominate the euclidean QCD
partition function. This can be achieved by restricting the (large) 
space-time volume $V$ to obey the inequalities 
\be
\frac 1{\Lambda_{\mbox{\rm{\small QCD}}}} \ll V^{1/4} \ll \frac 1{m_{\pi}},
\label{range}
\ee
where $m_{\pi}$ is the mass of the Goldstone bosons associated with the 
lightest quark mass  \cite{GL,LS}. 
The first inequality is required to separate
the Goldstone modes from the other hadronic excitations with a mass scale
of $\Lambda_{\mbox{\rm{\rm \small QCD}}}$ or higher.
When $V \to \infty$ the second inequality
gives an unphysical limit in which the pion is ultra-light, and never 
fits into the space-time volume. However, a wealth of information about
the Dirac operator spectrum can be computed {\em exactly} in this limit.

Once it is assured that only zero-momentum modes contribute to the
effective QCD partition function, its evaluation becomes remarkably
simple. The space-time integration of the effective Lagrangian density
yields just an overall volume factor, and the partition function becomes
identical to a zero-dimensional group integral over the coset determined by
the pattern of chiral symmetry breaking.  As was noted by Leutwyler
and Smilga \cite{LS}, it is highly advantageous to consider these partition 
functions in sectors corresponding to definite topological charge $\nu$ (in 
order to avoid absolute value signs, we 
will take $\nu$ positive or zero from now on).
Consider $SU(N_c\geq 3)$ gauge theories with $N_f$ fermions in
the fundamental representation. If chiral symmetry breaks according to
$SU_L(N_f)\times SU_R(N_f)\to SU_V(N_f)$, the effective partition function
of the zero momentum modes
takes on the form \cite{GL,LS}
\be
Z^{\rm GLS}_{N_f,\nu} ~=~ 
\int_{U(N_f)}\! dU ~{\det}^{\nu}(U) \exp\left \{V\Sigma_0~ {\mbox{\rm
Re\, Tr}}[{\cal M}U^{\dagger}]\right \} 
\label{ZLS}
\ee
in a sector of topological charge $\nu$, and with fermion mass matrix
${\cal M}$. Note that the effective partition function only depends on
masses $m_i$ and four-volume $V$ in the scaling combination of $\mu_i
\equiv m_iV\Sigma_0$. A few years ago it was
realized that this effective partition function itself has an entirely 
different representation in terms of large-$N$ Random Matrix Theory 
\cite{SVR,HVeff}. The precise relationship is as 
follows.  

Define a ``chiral'' random matrix partition function
with the same global symmetries as the field theory partition function
by {\cite{SVR,V}}
\be
Z_{N_f,\nu}^\beta(m_1,\cdots, m_{N_f}) =
\int DW \prod_{f= 1}^{N_f} \det({\rm \cal D} +m_f)
e^{-\frac{N \beta}4 {\rm Tr}V(W^\dagger W) },
\label{ranpart}
\ee
where
\be
{\cal D} = \left (\begin{array}{cc} 0 & iW\\
iW^\dagger & 0 \end{array} \right ),
\label{diracop}
\ee
and $W$ is a $n\times m$ matrix with $\nu = |n-m|$ and
$N= n+m$. The Random Matrix Theory potential $V(W^\dagger W)$ is not
restricted by any symmetry conditions, but it can be shown that all
relevant results in the proper limit do not depend on this potential once one
imposes that the {\em matrix} spectral density at the origin $\rho(0)$
be non-vanishing \cite{ADMN,brezin,Sener1,GWu,Seneru,Senerpr}.
The equivalent of the topological charge $\nu$  in the effective
partition function (\ref{ZLS}) is taken to be fixed. In the large-$N$ limit
we thus have $n = N/2$.
The matrix elements of $W$ are either real ($\beta = 1$, chiral
Gaussian Orthogonal Ensemble (chGOE)), complex
($\beta = 2$, chiral Gaussian Unitary Ensemble (chGUE)),
or quaternion real ($\beta = 4$, chiral Gaussian Symplectic Ensemble (chGSE))
\cite{V}.
For Yang-Mills theory with three or more colors and quarks in the fundamental 
representation the matrix elements of the Dirac operator are complex, and we 
have $\beta = 2$.
It can be demonstrated that in the microscopic domain (\ref{range}) 
the random matrix
partition function for any potential $V(W^{\dagger} W)$ can be mapped {\em
exactly} onto the effective finite volume partition function (\ref{ZLS}).
This was shown in ref. \cite{SVR,HVeff} for a
Gaussian potential, and, because the random matrix partition
function does not depend on $V(W^{\dagger} W)$ in this limit \cite{dam-review},
it holds for any potential. For more discussion of the random matrix partition 
function we refer to \cite{camreview}. 

Because of the spontaneous breaking of chiral symmetry, the
smallest eigenvalues of the Dirac operator are spaced as $1/\rho(0) = 
\pi/\Sigma_0 V$. In order to study the behavior of the smallest eigenvalues
in the approach to the thermodynamic limit, 
it is natural to rescale the eigenvalues
according to $u = \lambda V \Sigma_0$ and to introduce the microscopic limit
of the spectral density \cite{SVR}
\be
\rho_s(u) = \lim_{V\rightarrow \infty} \frac 1{V\Sigma_0} 
\rho(\frac u{V\Sigma_0}).
\label{rhosu}
\ee
For broken chiral symmetry this results in a nontrivial limiting function.

To understand the significance of the identification between the chiral
Lagrangian (\ref{ZLS}) and the Random Matrix Theory (\ref{ranpart}) in the
microscopic domain, it
helps to view these partition functions $Z_{\nu}[\mu_i]$
as generating functions for the chiral condensates $\langle\bar{\psi}_i
\psi_i\rangle$. Conventionally they are  defined by taking the zero-mass
limit in the end. However, it is useful to focus instead on the
mass-dependent chiral condensate, defined in the obvious way by
\be
\Sigma_i(\mu_1,\ldots,\mu_{N_{f}}) ~\equiv~ \frac 1V\frac{\partial}{\partial
\mu_i} \log Z_{\nu}[\mu_1,\ldots,\mu_{N_{f}}] ~.
\label{massdep0}
\ee  
As is evident from the spectral
representation of the condensate, this quantity carries much information 
about the microscopic spectral density of the Dirac operator: 
\be
\Sigma_i(\mu_1,\ldots,\mu_{N_{f}}) ~=~ \Sigma_0 \int du 
\frac{\rho_s(u;\mu_1,\ldots,\mu_{N_{f}}) }{iu + \mu_i}
\label{sigmarho}
\ee
The issue at hand is whether this relation can be uniquely inverted
to provide the microscopic spectral density 
$\rho_s(u;\mu_1,\ldots,\mu_{N_{f}})$ in terms of
the mass-dependent chiral condensate $\Sigma_i(\mu_1,\ldots,\mu_{N_{f}})$.
While the relation (\ref{sigmarho}) resembles the Stieltjes transform
of $\rho_s(u;\mu_1,\ldots,\mu_{N_{f}})$ (which under suitable convergence
criteria has a unique inverse), the $\mu_i$-dependence of the microscopic
spectral density ruins this identification. This problem was recently
solved in a paper by three of us \cite{OTV}, where
it was noted that the introduction of an additional fermion species into
the theory can be used to provide the needed unique inverse of a relation
of the form (\ref{sigmarho}). Clearly what is needed is that the 
microscopic spectral density becomes insensitive to the addition of this
additional species. The solution is to simultaneously introduce yet
another quark species, but this time of opposite statistics 
\cite{OTV}. In the original field theory
formulation this corresponds to a Euclidean partition function of the form
\be
Z_{\nu}^{\rm pq}  ~=~ \left(\prod_{f=1}^{N_{f}} m_f^{\nu}\right)\left(
\frac{m_{v1}}{m_{v2}}\right)^{\nu}
\int\! [dA]_{\nu} 
~\frac{\det(i\Dirac - m_{v1})}{\det(i\Dirac - m_{v2})}\prod_{f=1}^{N_{f}}
\det(i\Dirac - m_f) ~e^{-S_{YM}[A]} ~.
\label{pqQCD}
\ee
When $m_{v1}=m_{v2}$ this partition function simply coincides with the original
one. However, it is now also the generator of a mass-dependent chiral
condensate for the additional (say, fermionic) quark species, i.e.,
\be  
\Sigma(m_{v};m_1,\cdots, m_{N_f}) = 
\frac 1V \left .\frac {\partial}{\partial m_{v1}}
\right |_{m_{v1} = m_{v2}= m_v} \log Z_\nu^{\rm pq}.
\ee
In terms of the spectral density the valence quark mass dependence
of the chiral condensate can be rewritten as
\cite{Christ,vPLB,Trento}
\be  
\Sigma(m_{v};m_1,\cdots, m_{N_f})& =&  
\frac 1V \sum_k \left \langle \frac 1{i \lambda_k + m_v} 
\right \rangle \nonumber \\
&=&
\frac 1V
\int d\lambda \frac{\rho(\lambda; m_1,\cdots, m_{N_f})}
{i\lambda + m_v}.
\label{massdep}
\ee
Here,  $\langle \cdots \rangle$ denotes an average with respect to
the distribution of the eigenvalues.
Notice that the spectral density in this equation is for
$m_{v1}=m_{v2}$ and coincides with the original QCD one. 
Contrary to (\ref{massdep0}), the relation (\ref{massdep})
can then be inverted to give $\rho(\la;m_1,\ldots,m_{N_{f}})$.
As mentioned in \cite{OTV}, the spectral density 
follows from the discontinuity across the imaginary axis,
\be
\left .{\rm Disc}\right |_{m_v = i\lambda}\Sigma(m_v) 
= \lim_{\epsilon \rightarrow 0}
\Sigma(i\lambda+\epsilon) - \Sigma(i\lambda-\epsilon) = 2\pi \sum_k 
\langle \delta(\lambda +\lambda_k)\rangle
= 2\pi \rho(\lambda),
\label{spectdisc}
\ee
where we have suppressed the dependence on the sea-quark masses. 

The formulation of the spectral density  as a 
derivative of the partition function
(\ref{pqQCD}) provides a natural explanation for the result that 
the microscopic spectral density can be related to the usual finite
volume partition function with two additional flavors
\cite{Dampart,OTV}.

There is a close analogy to the quenching prescription of lattice
gauge theory, but it is worthwhile stressing that, since we eventually 
restrict ourselves
to equal masses $m_{v1}=m_{v2}$, there are {\em no} approximations
involved in introducing two additional quark species of opposite statistics
in this manner. 

The effective Lagrangian corresponding to the partition function
(\ref{pqQCD})  with two additional quark species is determined by its
underlying global (super-)symmetry structure. 
Its precise form is very close to the usual one, except for the fact that
the group manifold of the Goldstone modes is that of a supergroup.
The main object of this paper
is the calculation of the valence quark mass dependence 
of the chiral condensate from this effective theory. Below we will focus
on the domain (\ref{range}) where the nonzero momentum modes
factorize from the effective partition function. 
This super-symmetric
effective Lagrangian has been considered in a different
context by Bernard and Golterman \cite{Morel,qChPT}, who, 
by analogy with
lattice gauge theory, call it the  partially quenched chiral Lagrangian. 
The terms ``sea quarks'' for the original physical fermions
and ``valence quarks'' for the additional (fermionic)
species have been borrowed from lattice gauge theory as well.
In perturbation theory the cancellation of
the associated determinants in the path integral corresponds to the
omission of loops with this fermion. We stress again that in the present
context this is not an approximation, but precisely what is required for the
associated spectral density to be equal to the QCD spectral density.
The quenched version of this effective Lagrangian was recently derived
from a two-sublattice random flux model \cite{Simons-Altland} 
using the flavor-color
transformation introduced by Zirnbauer \cite{color-flavor}.

Below we will find that the microscopic spectral density 
obtained from the extreme infrared sector of
(\ref{pqQCD}) agrees with  the result \cite{V,VZ}
derived from chiral Random Matrix Theory. 
One can wonder how $exact$ results can be obtained this way.
The answer is simple.
In the microscopic scaling regime (\ref{range}) the effective QCD Lagrangian
is entirely free of dynamics. The only assumption is that the theory
supports spontaneous breaking of chiral symmetry. With this
knowledge alone the microscopic limit of the QCD partition function can
be written in terms of a zero-dimensional group integral over the
Goldstone manifold. This is both true for the usual chiral Lagrangian
and for the chiral Lagrangian corresponding to a partition function
with valence quarks. The latter results in the exact distribution of
the  eigenvalues of the Dirac operator in the microscopic scaling limit. 
This computable and {\em universal} end of the Dirac operator spectrum
is, as was to be expected, entirely independent of the detailed QCD dynamics.

The organization of this paper is as follows.
In section 2 we introduce the effective theory corresponding to
QCD with additional quark species
of different statistics and discuss the geometry of the Goldstone manifold.
Some of the pitfalls 
associated with the evaluation of super-integrals are discussed in section 3.
In section 4 we calculate the measure using an explicit representation of
the super-unitary group. 
In section 5 we evaluate the valence quark mass
dependence of the chiral condensate in the sector of topological charge $\nu$,
and derive directly from this the microscopic spectral density
$\rho_s(\la)$, which
is found to agree exactly with earlier results based on Random Matrix
Theory. A general expression for arbitrary number of flavors 
and topological charge is obtained in section 6. This expression is evaluated
in section 7 for an arbitrary number of massless quarks in the sector of
zero topological charge and for one sea quark in a sector of topological
charge $\nu$.
Concluding remarks are made in section 8. Additional technical details of the 
calculations can be found in appendices A and B.

\vskip1.5cm
\noindent
{\bf 2. Low Energy limit of QCD}
\vskip 0.5cm

If chiral invariance is spontaneously broken in QCD, the low energy limit 
of the theory is dominated by the Goldstone modes associated with this 
spontaneous symmetry breaking.
The basic flavor symmetry of the QCD $action$ with $N_f$ 
(physical) sea quarks and $N_v$ valence quarks is
\be
Gl_L(N_f+N_v|N_v) \otimes Gl_R(N_f+N_v|N_v).
\label{full-group}
\ee
The reason for this bigger symmetry group is that a priori we do not relate the
integration variables in the partition function by complex conjugation. 
This symmetry group is not necessarily a symmetry of the QCD {\it partition
function}. The symmetry transformations may violate the convergence of
the integrals. There are no problems for the Grassmann integration. However,
the integrations of the bosonic quark fields are only convergent if the
fields are related by complex conjugation. For this reason a $U_A(N_v)$ 
transformation on the $bosonic$ quark fields is not a symmetry of the
partition function. However, a $Gl(N_v)/U(N_v)$ axial transformation is 
consistent with convergence requirements.

 After
spontaneous breaking of the chiral symmetry according to
\be
Sl_L(N_f+N_v|N_v) \otimes Sl_R(N_f+N_v|N_v) \rightarrow 
Sl_V(N_f+N_v|N_v),\label{susybreaking}
\ee
the symmetry of the QCD partition function is reduced to 
$Sl_V(N_f+N_v|N_v)\oslash Gl_V(1)$ where $\oslash$ denotes the semi-direct
product. Notice that  
an axial $Gl(1)$ subgroup of the group (\ref{full-group}) is broken explicitly 
by the anomaly. 

The Goldstone manifold corresponding to the symmetry breaking pattern
(\ref{susybreaking}) is based on
the symmetric superspace $Sl_A(N_f+N_v|N_v)$. In our effective 
partition function the terms that break the axial symmetry will be
included explicitly resulting in an integration manifold given by 
$Sl_A(N_f+N_v|N_v) \otimes Gl_A(1)$. However, this manifold is
not a super-Riemannian manifold and is not suitable as an integration domain
for the low energy partition function. As an integration domain we choose
a maximum Riemannian submanifold of $Gl_A(N_f+N_v|N_v)$.
This results in a  fermion-fermion
block given by the compact domain $SU_A(N_f+N_v)$, 
whereas the boson-boson
block is restricted to the non-compact domain $Gl(N_v)/U(N_v)$. Because of
the super-trace, this compact/non-compact structure is required for obtaining
a positive definite quadratic form for the mass term and the
kinetic term of our low energy effective partition function
\cite{ChPTfound}. For a detailed
mathematical discussion of this construction we refer to a paper by
Zirnbauer \cite{zirnall}.

In this paper we restrict ourselves to the case of just one valence quark, 
$N_v =1$, which is all that is needed to derive the microscopic spectral
density of the QCD Dirac operator. 
We will denote our integration manifold by $Gl(N_f+1|1)$. The fields are
parametrized by  
\be
 U={\rm exp} (i\sqrt 2 \Phi/F), 
\ee 
where $\Phi$ is a $(N_f+2) \times (N_f+2)$ superfield: 
\be 
\Phi=\left(
\begin{array}{ll}
\phi & \bar \chi \\ \chi & i\tilde{\phi}
\end{array}
\right)  
\ee 
and $F$ is the pion decay constant.
Here, $\phi$ is a $(N_f+1) \times (N_f+1)$ Hermitean matrix containing the
ordinary mesons made of quarks and antiquarks. The factor $i$ in  the 
field  $\tilde{\phi}$ provides us with a parametrization of $Gl(1)/U(1)$.
It represents a meson made out of two ghost quarks.
Finally, $\chi$ and $\bar \chi$ (not related by complex conjugation)
represent fermionic mesons consisting of a ghost-quark and an ordinary
anti-quark. They are the Goldstone fermions associated with the spontaneous
breaking of the chiral supersymmetry (\ref{susybreaking}). 
We choose a diagonal mass matrix with $N_f$ sea quark masses
resulting in the  $(N_f+2) \times (N_f+2)$ quark-mass matrix  
\be 
\hat{\cal M}={\rm diag} ({m_1, \dots,
m_{N_f}},m_v,m_v-J).  
\ee
The source $J$ acts as to lift the degeneracy between the valence quark
and its superpartner. We will eventually set $J=0$.

The axial symmetry of the QCD partition function in 
the sector of topological charge $\nu$ is broken explicitly by a mismatch
in the number of left-handed and right-handed modes. Under a global  axial 
transformation $U_A$, the integrand in (\ref{pqQCD})
is multiplied by a factor ${\rm Sdet}^\nu (U_A)$.
Taking into account the explicit breaking by the mass term as well,
the QCD partition function in the range (\ref{range}) reduces to
the effective low energy partition function
\cite{OTV}
\be
Z^\nu_{N_f}(\hat{\cal M}) = 
\int_{U\in Gl(N_f+1|1)} dU \,{\rm Sdet}^\nu (U) e^{ 
 V\frac{\Sigma_0}{2} \; {\rm Str} (\hat{\cal M} U+\hat{\cal
M} U^{-1})}.
\label{superpart}
\ee
As discussed before, the integration manifold is given by the maximum 
Riemannian submanifold for the symmetric superspace $Gl(N_f+1|1)$.
For a definition of the super-determinant, Sdet,  and the 
super-trace, Str, we refer
to the book by Efetov \cite{Efetovbook}.

A consistency check of this effective partition follows
by comparing (\ref{superpart}) with (\ref{ZLS}), which implies
\be
Z^\nu_{N_f}(\hat{\cal M})\left.\right|_{J=0} ~=~  
Z^{\rm GLS}_{N_f,\nu}({\cal M}) ~.
\label{reduc}
\ee
While this identity may appear quite trivial, only a careful analysis
of the superintegral involved on the left hand side of this equation
will ensure that it is fulfilled. This issue is connected with the appearance
of Efetov-Wegner terms, which we will discuss next.

\vskip1.5cm
\noindent
{\bf 3. A Simple Example of a Superintegral}
\vskip 0.5cm
In this section we  remind the reader of some of the problems that occur in 
integrations
over a supermanifold. In spite of the fact that Grassmann integrations
are always convergent, the actual integration over a supermanifold can be
quite an involved task. 
Even an apparently innocent change of variables may result in a subtle 
paradox \cite{Efetovbook,Wegner}. To illustrate this fact,
let us look at an example given by Zirnbauer \cite{Zirnander}. Consider
the supermatrix
\be
A=\left(
\begin{array}{cc}
a&\alpha \\ \beta&i b
\end{array}
\right), \label{Aex}
\ee
and the following Gaussian integral
\be
{\cal I}=\int da \; db \; d\alpha \; d\beta \; e^{-\frac{1}{2} {\rm Str} A^2}
=\int da \; db \; d\alpha \; d\beta \; e^{-\frac{1}{2} 
(a^2+b^2)-\alpha \beta}. \label{intex}
\ee
By explicit integration or by Wegner's theorem \cite{Wegner}, 
one knows that ${\cal I}=1$. After Zirnbauer,
let us change variables according to 
\be 
A \rightarrow A'=\left(
\begin{array}{cc}
s-\sigma \tau (s-it) & -\tau (s-it) \\
\sigma (s-it) & i t -\sigma \tau (s-it)
\end{array}
\right). \label{changevar}
\ee
The Berezinian of this transformation is easily computed to be:
\be
\left|\frac{ \partial (a,b,\alpha,\beta) }{\partial (s,t,\sigma,\tau) } 
\right|=\frac{1}{(s-it)^2}.
\ee
Therefore the integral (\ref{intex}) becomes
\be
{\cal I}=\int ds \; dt \; d\sigma \; d\tau \; \frac{e^{-\frac{1}{2} Str A'^2}}
{(s-it)^2}
=\int ds \; dt \; d\sigma \; d\tau\; \frac{e^{-\frac{1}{2} (s^2+t^2)}}
{(s-it)^2}. \label{intex2}
\ee
The change of variables (\ref{changevar}) has removed all the Grassmann 
variables from the integrand. One might naively conclude that
${\cal I}=0$, in flagrant contradiction with the result we derived before 
with the original set of variables! However, this conclusion 
is wrong: The Berezinian is singular at $s=t=0$, and the integral
over $s$ and $t$ in (\ref{intex2}) is  therefore not defined. 

An even more puzzling substitution is to keep the
fermionic variables $\alpha$ , $\beta$ of (\ref{Aex}), while 
using the eigenvalues of
$A$ as bosonic integration variables \cite{GuhrNucA}. 
They are easily computed to be:
\be \left\{
\begin{array}{lll}
\tilde{a}&=&a+\frac{\alpha \beta}{a-ib}\\
i \tilde{b}&=&ib+\frac{\alpha \beta}{a-ib}. 
\end{array}
\right.
\ee
The Berezinian of this change of coordinates is equal to one, and
the integral (\ref{intex}) therefore becomes
\be
{\cal I}=\int d\tilde{a} \; d\tilde{b} \; d\alpha \; d\beta \; 
e^{-\frac{1}{2} (\tilde{a}^2+\tilde{b}^2)}.
\ee
Again the Grassmann variables have disappeared from the integrand, 
and one could naively conclude that ${\cal I}=0$. This is not 
true. The substitution above generates boundary contributions, 
the so-called Efetov-Wegner terms \cite{Efetovbook,Wegner}.
In order to 
compensate for the nilpotent terms in the integration domain of the 
superintegral, the measure has to be modified accordingly. The
correct measure, including the Efetov-Wegner term, is known in 
this example \cite{GuhrNucA}.

This short discussion exemplifies the subtleties involved in the 
computation of a superintegral. In a superintegral, one has to be 
suspicious about any change of
variables. When bosonic coordinates are shifted by nilpotent terms, 
the path of integration may contain even functions of the Grassmann variables. 
This 
can result in the notorious Efetov-Wegner terms in the measure 
\cite{Efetovbook,Wegner,zirnall,Rothstein,Alfaro,ZirnL,weidip}. 
This is not an academic problem. In our case, if 
the integration of the $Gl(N_f+1|1)$ partition function
is performed
via a supersymmetric generalization of the Itzykson-Zuber 
integral \cite{Guhr91,Alfarom,GW}, 
Efetov-Wegner terms appear in the measure in a way very similar to the
second change of variables in the example above. 
However, as is evident from the example, the possible appearance of
Efetov-Wegner terms depends on the choice of parametrization. 
Below we introduce a parametrization without such anomalous terms for the
observables under consideration. 
For example we will find that the partition function is properly normalized 
without the need for anomalous terms.


\vskip1.5cm
\noindent
{\bf 4. Parametrization of $Gl(N_f+1|1)$ and calculation of the measure}

\vskip 0.5cm

In order to construct an explicit parametrization for $Gl(N_f+1|1)$ 
we remind the reader of a parametrization of the Riemannian
superspace associated with  $U(1|1)$ which appeared
earlier in the literature \cite{Guhr22}, 
\be
\left ( \begin{array}{cc} e^{i\theta} & 0 \\ 0 & e^{i \phi} \end{array} 
\right ) \exp
\left ( \begin{array}{cc}  0 & \alpha \\ \beta & 0\end {array} \right ).
\label{u11}
\ee
The usefulness of this parametrization lies in the 
factorization into an ordinary and
a Grassmannian factor. As was discussed in previous section
the Goldstone manifold is a Riemannian superspace that requires a 
parametrization in terms of compact and non-compact variables
\cite{Andreev,zirnall,Sener1}. More specifically, instead of the
parametrization in (\ref{u11}), the boson-boson block is given by
$Gl(1)/U(1)$, obtained
by the replacement $\exp i\phi \rightarrow \exp s$.
The generalization to $Gl(N_f +1 |1)$ is immediate:
\be
U = \left ( \begin{array}{cc} U_n & \vec 0 \\ {\vec 0}^T & e^s \end{array} 
\right ) \exp
\left ( \begin{array}{cccc}  0 & \cdots & 0 & \alpha_1 \\
                          \vdots & & \vdots & \vdots \\
                          0 & \cdots & 0 & \alpha_n \\
                       \beta_1& \cdots & \beta_n & 0
         \end{array} \right ),
\label{para}
\ee
where $\vec 0$ is a null-vector of length $n\equiv N_f+1$, $U_n$ is a unitary 
matrix and the $\{\alpha_i\}$ and  $\{\beta_i\}$ are Grassmann variables.
If this product is written as 
\be
U = U_o U_g,
\ee
we find that
\be
\delta U' \equiv U_o^{-1} dUU^{-1} U_o =  U_o^{-1}dU_o + dU_g U_g^{-1}.
\ee
Since the Jacobian of a similarity transformation is unity, the invariant 
measure of the super-Riemannian manifold is given by
\be
d[U] = B(U_o,U_g) DU_n ds \prod_{k=1}^n  d\beta_k d\alpha_k,
\ee
where $DU_n$ is the invariant measure of $U(N_f +1)$, and
the Berezinian is given by
\be
B= {\rm Sdet} \frac {\delta{ U'}}
{\delta U_n \delta s  \delta \alpha_1 \cdots \delta \beta_n }.
\ee 
The $n^2$ differentials $\delta U_n$ are defined by 
$\delta U_n \equiv  U_n^{-1}dU_n$. {}From 
the definition of the superdeterminant
we find that the Berezinian factorizes as follows
\be
B =  \frac {B_1}{ B_2},
\ee
with
\be
B_1 =
{\rm Sdet} \frac {\delta{ U'}}{\delta U_n \delta s \delta {U_g}_{1n+1} \cdots 
\delta {U_g}_{nn+1}\delta {U_g}_{n+1\, 1}\cdots \delta {U_g}_{n+1\, n} },\, 
\label{B1}
\ee
and
\be
B_2 =
\det \frac {\delta {U_g}_{1n+1} \cdots 
\delta {U_g}_{n\, n+1}\delta {U_g}_{n+1\, 1}\cdots \delta {U_g}_{n+1\, n} }
{\delta \alpha_1 \cdots \delta \alpha_n
\delta \beta_1 \cdots \delta \beta_n}.
\ee
Here, $\delta U_g \equiv dU_g U_g^{-1}$. By inspection one finds that
the matrix in $B_1$ contains a block of zeros. One then trivially
verifies that $B_1 = 1$. This implies  the factorization of the measure
in an ordinary unitary invariant measure and a Grassmannian 
factor. 

The calculation of the second determinant proceeds as follows. We first
rewrite $U_g$ as 
\be
U_g = 1 + \frac {\cosh x - 1}{x^2} G^2 + \frac {\sinh x}{x} G,
\label{Ug}
\ee
where $G$ denotes the exponent in $U_g= \exp G $ and 
\be
x^2 = \sum_{k=1}^{N_f+1} \beta_k \alpha_k.
\ee
Here and below, $\sinh x/ x$ and $\cosh x$ are meant as a formal power series 
in $x^2$. The expression for $U_g$
can be easily derived by means of the identity $G^3 = x^2 G$.
After calculating $\delta U_g =  dU_g U_g^{-1}$ one obtains the 
following explicit expression for $B_2$,
\be
B_2 = \det \left [ \frac{\sinh x}{x}\left ( \begin{array}{cc}
\delta_{kl} + a\, \beta_k \alpha_l & b\, \beta_k \beta_l \\
-b\, \alpha_k \alpha_l & \delta_{kl} - a\, \alpha_k \beta_l
\end{array} \right ) \right ],
\label{berdet}
\ee
where 
\be
a &=& -\frac 1{2x \sinh x} + \frac 1{2x^2} - \frac{\cosh x - 1}{2x^2}, 
\nonumber \\
b &=& -\frac 1{2x \sinh x} + \frac 1{2x^2} + \frac{\cosh x - 1}{2x^2}.
\ee
A general expression for the determinant with the structure of (\ref{berdet})
is given in Appendix A. 
The final result for $B$ turns out to be remarkably simple
\be
B= \cosh x \left ( \frac x {\sinh x} \right )^{2N_f + 3} .
\label{berezinian}
\ee
For $N_f = 0$ one immediately finds that $B = 1$ and, for $N_f = 1$, the result
is $B= 1 + \frac 13(\alpha_1\beta_1 +\alpha_2\beta_2)$. 

Instead of (\ref{para}) one could use
an alternative parametrization defined by 
\be
U = V \Lambda V^{-1},
\ee
where $\Lambda$ is a diagonal matrix with matrix elements
$\Lambda_{kk}\exp i\theta_k$, $k = 1,\cdots, N_f + 1$ and 
$\Lambda_{N_f+2\,N_f +2}=\exp s$. The matrix $V$ is a (compact)
super-unitary matrix. The Berezinian from the transformation from
the $U$-variables to the $V$ and $\Lambda$ variables is given by
\be
\frac {\prod_{k<l} |e^{i\theta_k} -e^{i\theta_l}|^2}
{\prod_k (e^s -e^{i\theta_k})(e^{-s} -e^{-i\theta_k})}.
\label{izpar}
\ee
This parametrization is particularly useful if the integrals are 
calculated by means of a supersymmetric generalization of the Itzykson-Zuber 
integral \cite{Guhr91,Alfarom,GW}. 
However, in this case the 
Efetov-Wegner terms may be nonvanishing and must be carefully analyzed.

\vskip1.5cm
\noindent
{\bf 5. The Microscopic Spectral Density in the Quenched Limit} 
\vskip 0.5cm

The valence quark mass dependence of the chiral condensate for $N_f = 0$ and
$\nu = 0$
was calculated in \cite{OTV} using the
parametrization (\ref{izpar}). It is straightforward to generalize this
computation to arbitrary topological charge. As a warm-up exercise for the
calculation for arbitrary $N_f$, we first calculate 
$\Sigma(m_v)$ in a sector of topological charge $\nu$ using the 
parametrization (\ref{para}) for $N_f = 0$.

For $N_f = 0$ the parametrization (\ref{para}) 
yields a measure that is simply flat (see eq. (\ref{berezinian})). 
In this genuinely quenched limit, the zero mode 
partition function (\ref{superpart})  can be written as\\
\be
Z^{\nu}_{N_f=0} (m_v,m_v-J)
&=&\int \frac{d\theta}{2\pi} \; ds \;  d\beta \;d\alpha \; 
e^{\nu (i \theta-s)} 
\nonumber \\
&\times& \exp \left [ \Sigma_0 V \; {\rm Str} 
\left ( \begin{array}{cc}
            m_v&0\\0&m_v-J
         \end{array}
\right ) \;
\left ( \begin{array}{cc}
(1 +\frac{1}{2} \alpha \beta )\cos \theta & 
 \alpha (e^{i \theta}-e^{-s})/2 \nonumber \\
 \beta (e^s-e^{-i \theta})/2
& (1-\frac{1}{2} \alpha \beta) \cosh s
\end{array}
\right )
\right ]. \nonumber \\
\ee
The normalization of this partition function, $Z^{\nu}_{N_f=0} (m_v,m_v)$,
for $m_v \ne 0$ follows
by simply expanding the Grassmann variables. Using the Wronskian identity
for the modified Bessel functions,
\be 
K_\nu(x)I_{\nu+1}(x) + I_\nu(x)K_{\nu+1}(x) = \frac 1x,
\label{wronskian}
\ee
 we find $Z^{\nu}_{N_f=0} (m_v,m_v)= 1$. This is in  agreement
with Wegner's theorem for super-unitary invariant integrals \cite{Wegner}.
Moreover, in our context it has an immediate interpretation in
terms of eq. (\ref{reduc}), whose right hand side in this $N_f=0$ case
simply reduces to a mass-independent constant (which conveniently can be chosen
as unity, as done here).

After differentiation with respect to $J$ and a 
trivial integration over the 
Grassmann variables, the valence quark
mass dependence of the chiral condensate  is found to be
\be
\Sigma(m_v)=\frac{\Sigma_0}{2} \int \frac{d\theta}{2\pi} \; ds \;
e^{\mu_v (\cos \theta-\cosh s)} \; e^{\nu (i \theta-s)} \; \Big[ \cosh s
(\mu_v\cos \theta + \mu_v \cosh s - 1)
  \Big],
\ee
where $\mu_v=m_v V \Sigma_0$. This integral is easily computed, it can
be expressed in terms of Bessel functions. Using standard recursion relations
and the Wronskian identity  (\ref{wronskian}) the result we get is:
\be
\frac{\Sigma(m_v)}{\Sigma_0} = \mu_v \Big[ I_\nu(\mu_v) K_\nu (\mu_v)
+I_{\nu+1}(\mu_v)
K_{\nu-1} (\mu_v) \Big] +\frac{ \nu}{\mu_v}.
\label{Nf0sigma}
\ee
The last term corresponds to the number of zero modes of 
the Dirac operator. It is remarkable that this term, which has such a simple
interpretation in the framework of the original QCD partition function
(where it comes from the explicit factor $m^{\nu}$ in eq. (\ref{pqQCD})),
is reproduced by the supersymmetric chiral Lagrangian (which does not 
contain such an explicit  factor $m^{\nu}$). This phenomenon is general,
and occurs in the framework of the usual chiral Lagrangian as well 
\cite{LS,msumrules}.

The result (\ref{Nf0sigma}) coincides exactly with the valence quark mass 
dependence of the chiral condensate obtained from Random Matrix Theory 
by integrating
the microscopic spectral density according to (\ref{massdep}) \cite{vPLB}. 
What is most important in the present context is that this relation can be 
inverted. Using eq. (\ref{spectdisc}) we immediately find
\be
\rho_s(u) ~=~ \frac{u}{2}\left[J_{\nu}(u)^2 - J_{\nu+1}(u)J_{\nu-1}(u)
\right] ~.
\label{rhosu0}
\ee
This is the microscopic spectral density of the Dirac operator in QCD for
$N_f=0$ in a sector of arbitrary topological charge $\nu$. It agrees
exactly with the result obtained earlier by means of chiral Random Matrix
Theory \cite{V}. Here,
it has been derived directly from the effective finite-volume partition
function of QCD in the microscopic scaling regime. 

The results derived in this section 
 can be tested by means of lattice QCD simulations.
Results obtained with staggered fermions for three colors
convincingly show that the valence
quark mass dependence of the chiral condensate \cite{vPLB}
and the microscopic
spectral density \cite{pandt} are given by the above expressions for 
$\nu =0$. Apparently,  effects of the topological charge become only
visible in  simulations close to continuum limit. For completeness
we also mention 
that lattice QCD results for the microscopic spectral
density for $N_c= 2$ with staggered
fermions is given by similar universal expressions as well
\cite{Tilo}.

\vskip 1.5cm
\noindent
{\bf 6. The Microscopic Spectral Density for QCD with $N_f$ Flavors} 
\vskip 0.5cm

Let us now turn to the general case of $N_f$ flavors and topological 
charge $\nu$.
The (partially) supersymmetric effective partition function
is given in eq. (\ref{superpart}). It will turn out that
the measure of this partition function is normalized such that 
$Z^\nu_{N_f}({\cal M}) = N_f!$ for $ J = 0$ and zero sea quark masses. 
In general,
it must satisfy the identity (\ref{reduc}) up to a normalization factor. 

We compute the valence quark mass dependence of the chiral condensate by
\be
\Sigma(m_v) = \frac 1V \left . \partial_J 
\log Z^\nu_{N_f}(\hat{\cal M})\right 
|_{J= 0} ~.
\ee
Using the  parameterization (\ref{para}) with $U = U_o U_g$, 
we can rewrite the partition function as
\be
Z^\nu_{N_f}(\hat{\cal M}) = \int d[U] {\rm Sdet}^\nu (U)\exp \left\{ 
\frac{\Sigma_0 V}2
{\rm Str}[ U_g\hat{\cal M} U_o + \hat{\cal M }U_g^{-1} U_o^{-1}] \right\}.
\ee
Separating out the fermion-fermion ($FF$) and the boson-boson ($BB$) blocks 
we have in the exponential
\be
Z^\nu_{N_f}(\hat{\cal M}) = \int d[U] {\rm Sdet}^\nu (U)\exp 
\left \{ {\Sigma_0 V}\left ( \frac 12 {\rm Tr}[ U_g^{FF} \hat{\cal M}_n U_n + 
\hat{\cal M}_n U_g^{FF} U_n^{ -1}]
- \hat{\cal M}^{BB} U_g^{BB} \cosh s \right )\right\},
\label{e49}
\nonumber\\
\ee
where the $FF$ superscript stands for the upper left $(N_f+1)\times(N_f+1)$ 
block and $BB$ represents
the lower right element. The $FF$ blocks of $\hat{\cal M}$ and $ U$ are denoted
by $\hat{\cal M}_n$ and $U_n$, respectively. We have also used the property that 
$U_g$ and $U_g^{-1}$, have the same $FF$ blocks and $BB$ blocks
which is easily seen from the definition
of $U_g$ and its expansion in powers of the  Grassmann elements.

Using that ${\rm Sdet}^\nu (U) = \exp(-\nu s) {\det}^\nu(U_n)$ 
the $s$ integral can be done immediately resulting 
in  a modified Bessel function with argument
$\hat{\cal M}^{BB} U_g^{BB} = \mu_J \cosh x$ where $\mu_J$ is
defined as $\mu_J \equiv (m_v -J)V\Sigma_0$.
The next step in evaluating the partition function is to expand the exponent in
terms of the Grassmann variables in $U_g^{FF} = 1 + G^2(\cosh x - 1)/x^2  $
(see eq. (\ref{Ug})). Exploiting the rotational invariance of the 
scalar product $x^2$, the exponent can be expressed in terms of the 
eigenvalues, $\Lambda_k$
of the matrix  
\be
A = \Sigma_0 V (\hat{\cal M}_n U_n + U_n^{-1} \hat{\cal M}_n)/2. 
\ee
This allows
us to make the following replacement in the integrand of (\ref{e49}) 
\be
\exp \left \{   
{\rm  Tr} \,A U_g^{FF}  \right\} 
\rightarrow  \exp \left \{ 
  \sum_k \Lambda_k(1 + \frac{\cosh x - 1}{x^2}\
 \alpha_k \beta_k) \right \}. \nonumber \\
\ee 
Writing the exponent of the Grassmann variables as a product and once more
using the symmetry of the Grassmann variables in the integral we find
that an even  more symmetric integrand is obtained by making the
replacement
\be
\exp \left \{ 
 \frac{\cosh x - 1}{x^2}
\sum_k \alpha_k \beta_k \Lambda_k \right \} &\rightarrow&
\sum_{l=0}^{n} \left( {1-\cosh x} \right)^l
\frac{(n-l)!}{n!} \sum_{1 \le k_1 < \cdots < k_l \le n} 
\Lambda_{k_1} \cdots \Lambda_{k_l} \nonumber \\
&=&
\sum_{l=0}^{n} \left( {1-\cosh x} \right)^l
\frac{(n-l)!}{n!} \left .
\frac {(\partial_y)^l}{l!}\right |_{y=0}
\exp  {\rm Tr} \log (1+yA). \nonumber \\
\ee
The partition function can thus be written as
\be
Z^\nu_{N_f}(\hat{\cal M}) = 2\int \prod_k d\beta_k d\alpha_k 
\cosh x \left ( \frac x {\sinh x} \right )^{2n + 1} 
K_\nu(\mu_J\cosh x)
\sum_{l=0}^{n} \left( {1-\cosh x}\right)^l
\frac{(n-l)!}{n!} \Omega_l(\hat{\cal M}_n), \nonumber \\
\label{withber}
\ee
where
\be
\Omega_l(\hat{\cal M}_n )= \int dU_n {\det U_n}^\nu 
\left . 
\frac{(\del_y)^l}{l!} \right |_{y=0} \det (1+y A)
\exp \left \{ {\rm Tr}A \right\}.
\ee
In (\ref{withber}) we have included our result for the Berezinian 
(\ref{berezinian}) which factorizes into an ordinary part an a Grassmannian
part.

Up to a factor $(-1)^n n!$, 
performing the final Grassmann integrals is equivalent to finding the 
coefficient of $x^{2n}$ in the remaining integrand.  Using the residue theorem
the latter may be expressed as a contour integral around the pole at $x=0$
of the quotient of the integrand and $x^{2n+1}$.
We may then make the substitution $z=\sinh x$ to simplify the integrand
resulting in
\be
Z^\nu_{N_f}(\hat{\cal M}) = 2\oint \frac {dz}{2\pi i} \frac {(-1)^n n!}
{z^{2n+1}}  
K_\nu(\mu_J\sqrt{1+ z^2})
\sum_{l=0}^{n} \left( 1-\sqrt{1+z^2} \right)^l
\frac{(n-l)!}{n!} \Omega_l(\hat{\cal M}_n).
\ee

The term containing $K_\nu$ can be expanded using the product theorem for
Bessel functions which can be written as
\be
 K_\nu(\mu_J\sqrt{1+z^2} ) = (1+z^2)^{\nu/2}
\sum_{k=0}^{\infty} \frac1{k!} K_{k+\nu}(\mu_J) 
\left[ \frac{-z^2}{2} \mu_J \right]^k.
\ee
The other factor in the integrand can also be expanded in a series in $z^2$.
After multiplying the two out and extracting the coefficient of $z^{2n}$
we are left with the the final result for the partition function
\be
Z^\nu_{N_f}(\hat{\cal M}) = 
2(-1)^n\sum_{k=0}^n \sum_{l=0}^n \sum_{s=0}^l (-1)^s
\left(  \begin{array}{c} l \\s \end{array} \right )
\left(\begin{array}{c} \frac{s+\nu}2 \\ n-k  \end{array} \right )
\frac 1{k!} \left ( \frac {-\mu_J}2 \right )^k K_{k+\nu} (\mu_J) 
{(n-l)!} \Omega_l(\hat{\cal M}_n).\nonumber \\
\label{final}
\ee
For $\nu = 0$ this result can be further simplified by using the combinatorial
identity
\be
\sum_{s=0}^l (-1)^s\left( \begin{array}{c} l\\s \end{array} \right)
\left (\begin{array}{c} \frac s2 \\ p \end{array} \right) =
\left(\frac{-1}{4}\right)^p l 2^l \frac{(2p-l-1)!}{(p-l)!p!}
\ee
for $0 \le l \le p$ and it gives zero otherwise.  Here we take the right hand
side of this expression to be one when $p=l=0$.

        In general, the integrals over the unitary group give a fairly 
complicated expression.
However, the calculation greatly simplifies for an arbitrary number of 
massless quarks and zero topological charge. 
Of course, the calculation is relatively simple for the case 
of $N_f = 1$, even with the quark masses included and arbitrary
topological charge. 

\vskip 1.5 cm
\noindent
{\bf 7. Some Special Cases} 
\vskip 0.5 cm

While the above formulas are quite involved in general, there are some
important special cases in which they can be reduced to simple expressions
in terms of elementary functions.
Consider first the case of one physical quark of mass $m_1$, $i.e.$
$N_f =1$. The expression (\ref{final}) 
then reduces to
\be
Z^{\nu}_{N_f=1}(\hat{\cal M}) &=&
\frac12 K_\nu(\mu_J) [ \nu(\nu-2) \Omega_0 - (\nu - \frac12 ) \Omega_1 
                       + \Omega_2 ]
-\frac12 \mu_J K_{1+\nu}(\mu_J) [ 2 \nu \Omega_0 - \Omega_1 ]
\nonumber \\
&+&\frac12 \mu_J^2 K_{2+\nu}(\mu_J) \Omega_0.
\ee
Using the  $\Omega_l$ as computed in appendix B, we find that the
normalization of the partition function is given by
\be
\left . Z^{\nu}_{N_f=1}(\hat{\cal M})\right |_{J=0} =  I_\nu(m_1).
\ee
This result was to be expected. It coincides
with the finite volume partition function for $N_f = 1$
as derived in \cite{LS} (without valence quarks) and is just a
special case of (\ref{reduc}).  
The valence quark mass dependence of the chiral condensate follows by
differentiating $\log Z^\nu_{N_f}(\hat{\cal M})$
with  respect to the source term $J$.
This results in
\be
\frac{\Sigma(m_v;m_1)}{\Sigma_0} &=&
{\mu_v} \left[ I_{\nu+1}(\mu_v) K_{\nu+1}(\mu_v)
           + I_{\nu+2}(\mu_v) K_\nu(\mu_v)     \right]
+ \frac{\nu}{\mu_v} \cr && 
+2\mu_1 \frac{K_\nu(\mu_v)}{I_\nu(\mu_1)}
\frac{ \mu_v I_\nu(\mu_v) I_{\nu+1}(\mu_1)-\mu_1 I_\nu(\mu_1) I_{\nu+1}(\mu_v)}
{\mu_v^2-\mu_1^2} ~.
\ee
This can easily be compared with results computed from Random Matrix Theory
\cite{DN,WGW} by integrating the microscopic spectral density obtained
there according to eq. (\ref{massdep}). However, what is important here is
not the precise form of this mass dependent chiral condensate, but again
the fact that the relation (\ref{massdep}) can now be {\em inverted} to 
yield the microscopic spectral density itself. Using eq. (\ref{spectdisc})
we find
\be
\rho_s(u;m_1) = \frac{u}{2}\left[J_{\nu+1}(u)^2-J_{\nu+2}(u)J_\nu(u)\right] 
+ \mu_1 \frac{J_\nu(u)}{I_\nu(\mu_1)}
\frac{\mu_1 I_\nu(\mu_1)J_{\nu+1}(u) - 
uI_{\nu+1}(\mu_1)J_\nu(u)}{(u^2+\mu^2_1)} ~.
\ee
This result agrees exactly with the answer obtained from Random Matrix
Theory \cite{DN,WGW}.
 
One can of course continue by considering a larger number of massive flavors.
But the computations rapidly get rather involved, and we have not succeeded
in finding a simple compact expression for the mass dependent chiral
condensate for the general case of $N_f$ massive flavors in a sector
of arbitrary topological charge $\nu$. There is, however, one
important case which can be computed rather easily: that of an arbitrary number
of massless quarks in a sector of zero topological charge. In this case 
 the calculation greatly simplifies since then $\Omega_l=0$ for 
$l \ge 3$.  This is
readily seen from its definition in terms of the eigenvalues of $A$ and the 
observation that for massless sea-quarks only two eigenvalues of $A$ are 
nonzero.
In this case the expression (\ref{final}) reduces to
\be
Z^{\nu=0}_{N_f}(\hat{\cal M}) = 2\left ( \frac {\mu_J}2 \right )^{n}
 K_n(\mu_J){\Omega_0} + 
\left ( \frac {\mu_J}2 \right )^{n-1} K_{n-1}(\mu_J) {\Omega_1},
\ee
where we have used the identity (see appendix B)
\be
(n-1) \Omega_1  + 2\Omega_2 = 0,
\ee
which is valid for the case of zero sea quark masses and topological
charge equal to zero. 
{}From the explicit expressions in Appendix B for $\Omega_0$ and $\Omega_1$
we obtain our final result for the partition function
\be
Z^{\nu=0}_{N_f}(\hat{\cal M}) = (n-1)!\left [
\mu_J^n K_n(\mu_J) \frac{I_{n-1}(m_v)}{m_v^{n-1}}
+ \mu_J^{n-1} K_{n-1}(\mu_J) \frac {I_n(m_v)}{m_v^{n-2}}\right ]. \nonumber
\\
\ee
Next, from the Wronskian identity (\ref{wronskian}) we immediately find that
$Z(\hat{\cal M}) = (n-1)!= N_f!$ for $J=0$. 
The valence quark mass dependence of the 
chiral condensate follows from differentiation with respect to $J$, resulting
in
\be
\frac{\Sigma(m_v)}{\Sigma_0 }=  {\mu_v} \left [ I_{N_f}(\mu_v) K_{N_f}(\mu_v) 
+ I_{N_f+1}(\mu_v) K_{N_f-1}(\mu_v) \right ] ~.
\label{vsigma}
\ee
This expression agrees with the result obtained previously from Random
Matrix Theory \cite{vPLB}. Again our purpose here is just the opposite: we
can now {\em derive} directly from field theory the microscopic spectral 
density of the Dirac operator from the discontinuity of $\Sigma(m_v)$
as given in eq. (\ref{vsigma}). The result is
\be
\rho_s(u) ~=~ \frac{u}{2}\left [J_{N_{f}}(u)^2 - J_{N_{f}+1}(u)J_{N_{f}-1}(u)
\right ] ~.
\ee
This is the celebrated result obtained first using Random Matrix Theory
\cite{VZ}. Indeed, this taken together with the earlier results constitute 
an analytical proof that the 
smallest eigenvalues of the QCD Dirac operator are correlated according to 
a Random Matrix Theory whose form is dictated by the global symmetries of the 
QCD Dirac operator.

\vskip 1.5cm
\noindent
{\bf 8. Conclusions}
\vskip 0.5cm

In the limit of vanishing light quark masses the infrared sector of the 
QCD partition function is dominated by the Goldstone modes associated with 
the assumed spontaneous breaking of chiral symmetry. However, the usual 
chiral Lagrangian does not allow us to access the Dirac spectrum. For that
reason one has to extend the partition function with one valence quark and
its superpartner \cite{OTV}. The chiral Lagrangian of this partition function
is based on the super-group $Gl(N_f+1|1)\times Gl(N_f+1|1)$.
In agreement with a supersymmetric generalization of the Vafa-Witten theorem
\cite{Vafa} 
and the maximum breaking of chiral symmetry, the symmetry is broken
to the diagonal subgroup $Gl(N_f+1|1)$. As dictated by the convergence of the
integrals the integration manifold is restricted to a super-Riemannian 
symmetric submanifold. It is characterized by a symbiosis between 
compact and non-compact degrees of freedom. Remarkably, precisely this
balance between compact and non-compact variables in the effective
Lagrangian is what leads to the correct cut structure in the complex
valence quark mass plane. The supersymmetric extension of the effective
chiral Lagrangian has thus passed a highly non-trivial self-consistency
test. Moreover, in the present context this super-extension of the
chiral Lagrangian for QCD is not used to study artifacts
of (partially) quenched numerical simulations, but rather to derive
physical results in standard QCD with dynamical fermions.     

In a previous paper \cite{OTV}, three of us have shown that the microscopic 
spectral density can be obtained from a partition function with {\em compact} 
degrees of freedom only, i.e. by replacing $Gl(1)/U(1) \rightarrow U(1)$. 
The integrals over the supergroups could be performed conveniently by 
means of a supersymmetric generalization of the Itzykson-Zuber integral. 
However, a similar approach applied to the direct 
calculation of the valence quark mass 
dependence of the chiral condensate only worked for the quenched
approximation, i.e. for  the $Gl(1|1)$ partition function, but 
failed in the case of a nonzero number of sea quarks.
The general reason for such failure  is well understood. 
The measure contains anomalous terms which have to be included. 
The appearance of these so-called Efetov-Wegner terms has its
origin in  the fact that the  
integration contour contains nilpotent terms, which after expansion in a
power series, result in total derivatives. Typically, such terms contribute
in the neighborhood of singularities. 

In this work we have followed a different approach. We have directly 
calculated the $Gl(N_f+1|1)$ partition function without relying
on super-symmetric Itzykson-Zuber integrals. Technically, this
calculation was possible because we found a parametrization 
without anomalous contributions for the observables under consideration.

In conclusion, we have shown analytically that the microscopic distribution of
eigenvalues of the QCD Dirac operator can be computed directly from
a supersymmetric
extension of the effective finite-volume QCD partition function.
The results agree exactly with the original computations
which were based on chiral Random Matrix Theory. It is quite remarkable
that what could appear as a forbiddingly difficult field-theory
computation of the microscopic Dirac operator spectrum was {\em first}
performed on the basis of universality arguments and Random Matrix Theory.
What has now been proven is that these results are {\em exact} and indeed can 
be derived directly from field theory, without recourse to Random Matrix 
Theory. The underlying reason should be clear: Random Matrix Theories with
the global symmetries of the QCD partition function can be reduced to the
finite volume partition function that has been studied 
in this paper.


\vskip 1.5cm
\noindent
{\bf Acknowledgements}

\vskip 0.5cm

This work was partially supported by the US DOE grant
DE-FG-88ER40388. 
The work of P.H.D. was supported in part by EU TMR grant ERBFMRXCT97-0122,
 and the work of D.T. was supported by Schweizerischer Nationalfonds.
T. Guhr, A. Sch\"afer, A. Smilga, M. Stephanov, R. Szabo, H.A. Weidenm\"uller 
and T. Wettig are acknowledged for useful discussions. We particularly
benefitted from discussions with A. Altland and M. Zirnbauer 
on the super-unitary measure.

\vskip 1.5cm
\noindent
{\bf Appendix A. Calculation of a determinant}
\vskip 0.5cm

In this appendix we show that

\be
\det \left ( \begin{array}{cc}
  \delta_{kl} + a\, \beta_k \alpha_l &  b\, \beta_k \beta_l \\
-b\, \alpha_k \alpha_l & \delta_{kl} - a\, \alpha_k \beta_l
\end{array} \right ) = (1 - 2 a x^2 + (a^2-b^2) x^4)^{-1}.
\ee
Here, $\alpha_k$ and $\beta_k$ are Grassmann variables and $a$ and $b$
are scalar functions.
We use the notations that $\alpha_k \alpha_l$ represents an $n\times n$ matrix
with entries $\alpha_k \alpha_l$ and $x^2$ is defined by
\be
x^2 = \sum_{k=1}^n \beta_k \alpha_k.
\ee

The proof is as follows. If we introduce the matrix
\be
A = \left (\begin{array}{cc}
  a\, \beta_k \alpha_l &  b \,\beta_k \beta_l \\
-b\, \alpha_k \alpha_l&  - a\, \alpha_k \beta_l \end{array} \right )
\ee
the determinant is calculated by the relation
\be
\det (1+A) = \sum_{k=1}^\infty \exp \left [
\frac {(-1)^{k+1}}k {\rm Tr} A^k \right ].
\ee
One can easily show that the powers of $A$ have the same structure as
the matrix $A$. Therefore,
and
\be
A^p = \left (\begin{array}{cc}
  a_p\, \beta_k \alpha_l&  b_p \,\beta_k \beta_l \\
-b_p \,\alpha_k\alpha_l&  - a_p\, \alpha_k \beta_l \end{array} \right ).
\ee
It is straightforward to derive a recursion relation for the coefficients
$a_p$ and $b_p$
\be
a_{p+1} = -a x^2 a_p - bx^2 b_p,\\
b_{p+1} = -b x^2 a_p - a x^2b_p,
\ee
with $a_1 = 1$ and $b_1 = b$. The solution of these recursion relations is
given by
\be
a_p &=& \frac 12 (a+b)^p (-x^2)^{(p-1)} + \frac 12 (a-b)^p (-x^2)^{(p-1)}
\nonumber \\
b_p &=& \frac 12 (a+b)^p (-x^2)^{(p-1)} - \frac 12 (a-b)^p (-x^2)^{(p-1)}.
\ee
By resumming the power series into a logarithm and taking the trace of the
matrices, one easily recovers the expression for the determinant.

\newpage
\noindent
{\bf Appendix B. Some integrals over Unitary groups}
\vskip 0.5cm

In this appendix we calculate  the integrals
\be
\Omega_l(\hat{\cal M} )= \int dU {\det U}^\nu 
\left . 
\frac{(\del_y)^l}{l!} \right |_{y=0} \det (1+y A)
\exp \left \{ {\rm Tr}A \right\},
\label{unintdef}
\ee
where
\be
A = \Sigma_0 V (\hat{\cal M} U + U^{-1} \hat{\cal M})/2, 
\ee
and the integral is over the unitary group $U(n)$. We evaluate these
integrals for $l = 0, \, 1$ and 2.
These integrals are of the form
\be
\int d[U] f(U,U^\dagger) \det (U)^\nu
\exp \left\{ \frac12 {\rm Tr} [ M ( U + U^{\dagger} ) ] \right\} \equiv
\left< f(U,U^\dagger) \right>
\ee
with $M$ a diagonal matrix and $U$ ranging over the group ${ U}(n)$.
To evaluate these integrals we use the following result
\cite{browersu}
\be
W(J,J^\dagger) &=&
\int d[U] {\det(U)}^\nu
\exp \left\{ {\rm Tr} [ J U + U^{\dagger} J^{\dagger} ] \right\}
\nonumber\\
&=&
\prod_{k=1}^{n-1} 2^k k!
\left( \frac{\det J^\dagger}{\det J} \right)^{\frac\nu2}
\frac{{\det}_{ij}[z_i^{j-1} I_{j-1+\nu}(z_i)]}{\Delta(z_i^2)}
\ee
where $z_i=2 \sqrt{\lambda_i}$ with the
$\lambda_i$ being the eigenvalues of the matrix $J^{\dagger} J$ and
$\Delta(z_i^2)$ is the Vandermonde determinant $\prod_{k >l} (z_k^2 - z_l^2)$.
{}From the invariance of the measure and the fact that an arbitrary complex
matrix can be diagonalized by two unitary matrices resulting in a matrix
with positive diagonal elements, 
it follows immediately that the integral factorizes into  a function of the
eigenvalues of $J^\dagger J$ and a ratio of determinants of $J^\dagger$ and 
$J$. 
This result can be proven naturally \cite{JSV} by means of a generalization
of the Itzykson-Zuber integral to arbitrary complex matrices.
We can now rewrite the original integral as
\be
\left< f(U,U^\dagger) \right> =
f\left(\frac{\partial}{\partial J}^T,
\frac{\partial}{\partial J^\dagger}^T\right)
\left. W(J,J^\dagger) \right|_{J=J^\dagger=M/2}.
\ee

Since $W$ contains the eigenvalues $\lambda$, we will need to change
the derivatives with respect to the sources $J$ and $J^\dagger$ into 
derivatives on $\lambda$ by the chain rule
\be
\frac{\partial}{\partial J_{i j}} = \sum_{a=1}^n
\frac{\partial \lambda_a}{\partial J_{i j}}
\frac{\partial}{\partial \lambda_{a}}.
\ee
The derivative of $\lambda$ with respect to the sources can then be
obtained from the identities
\be
\sum_{p=1}^{n} \lambda_p^r = {\rm Tr} (J^{\dagger} J)^r
\label{trident}
\ee
for $r=1, \cdots, n$.
By applying derivatives with respect to the sources $J$ and $J^\dagger$
on both sides of
the equation we get a system of equations which we can then solve for the
required derivative of $\lambda$.

As an example we calculate the integrals corresponding to $\Omega_1$ and 
$\Omega_2$ as defined in (\ref{unintdef}) which are used in the text.
Our goal is to
express them in terms of $\Omega_0 = <1>$ and its derivatives with respect
to the masses.  We can then use the explicit expression for $\Omega_0$ for
the two cases considered in the text to obtain an expression for these
integrals.  We consider  the case  with  $N_f=1$ sea quark with mass $m_1$,
one valence quark with mass $m_v$ ($n=2$) for topological charge
$\nu$ 
and the case with 
$N_f$ massless flavors,  one valence quark with mass $m_v$ ($n=N_f+1$) 
for zero topological charge ($\nu=0$).  In these
cases, the values of $\Omega_0$ are given
by 
\be
\Omega_0 =
2 \frac{m_v I_{1+\nu}(m_v) I_{\nu}(m_1) - m_1 I_{1+\nu}(m_1) I_{\nu}(m_v)}
{m_v^2-m_1^2},
\ee
and
\be
\Omega_0 =  2^{N_f} N_f! \frac{I_{N_f}(m_v)}{m_v^{N_f}},
\ee
respectively. 
 For simplicity we absorb the factors of $\Sigma_0 V$ into the masses in this
appendix. We emphasize that in this appendix the sea quark masses and
the valence quark mass enter on the same footing. We nevertheless make
this distinction since it allows us to make contact
with the formulas in the main text. 

First we calculate
\be
\Omega_1 = <{\rm Tr} A> = \frac12 \sum_i^n m_i
\left( \frac{\partial}{\partial J_{i i}} +
\frac{\partial}{\partial J^{\dagger}_{i i}} \right)
\left. W(J,J^\dagger) \right|_{J=J^\dagger=M/2}.
\ee
One can check that the derivatives acting on the term
$(\det J^\dagger/\det J)^{\nu/2}$ gives no contribution to the final result.
Since this calculation only requires diagonal sources,
it is easy to see that the derivatives acting on the remaining term gives
\be
\Omega_1 = \sum_i^n m_i \frac{\partial}{\partial m_i} \Omega_0.
\ee
For the two cases above, for $N_f=1$ and one valence quark
$(n=2)$ in the sector of topological charge $\nu$ we obtain
\be
\Omega_1 = 2 I_{\nu}(m_v) I_{\nu}(m_1)
-2 \Omega_0,
\ee
and for $N_f$ massless flavors and one valence quark ($n=N_f+1$) 
with zero topological charge ($\nu = 0$) we find
\be
\Omega_1 = 2^{N_f} N_f! \frac{I_{N_f+1}(m_v)}{m_v^{N_f-1}}.
\ee

The next integral,
\be
\Omega_2 = \frac12 \left< ({\rm Tr} A)^2 - {\rm Tr} A^2 \right>,
\ee
includes off diagonal sources and thus becomes more complicated.
The first term is similar to the previous example and can be written
\be
\left< ({\rm Tr} A)^2 \right> = \sum_{i,j}^n m_i m_j 
\frac{\partial}{\partial m_i}
\frac{\partial}{\partial m_j} \Omega_0.
\ee
Here again the term $(\det J^\dagger/\det J)^{\nu/2}$ in $W$ did not
contribute.
The second term is
\be
\left< {\rm Tr} A^2 \right> =
\frac14 \left< {\rm Tr} [ 2 M^2 + (MU)^2 + (U^\dagger M)^2 ] \right>.
\ee
We thus need to calculate $< {\rm Tr} MUMU >$ and its conjugate.
In order to evaluate these integrals we write
\be
\left< {\rm Tr} (MU)^2 \right> = \sum_{i,j}^n m_i m_j
\frac{\partial}{\partial J_{i j}}
\frac{\partial}{\partial J_{j i}}
\left.  W(J,J^\dagger)\right|_{J=J^\dagger=M/2},
\ee
and similarly for its conjugate.  Now the term
$(\det J^\dagger/\det J)^{\nu/2}$ does make a contribution.  Summing the
contribution for this case and its conjugate gives $2n\nu^2\Omega_0$.
The remainder of these integrals can again be evaluated
using the chain rule
\be
\frac{\partial}{\partial J_{i j}}
\frac{\partial}{\partial J_{j i}} =
\sum_a^n \frac{\partial^2 \lambda_a}{\partial J_{i j} \partial J_{j i}}
\frac{\partial}{\partial \lambda_a}
+\sum_{a,b}^n \frac{\partial \lambda_a}{\partial J_{i j}}
\frac{\partial \lambda_b}{\partial J_{j i}}
\frac{\partial^2}{\partial \lambda_a \partial \lambda_b}.
\ee
Applying the two derivatives to the identities (\ref{trident})
and then solving the resulting
system of equations, we find (for $i \ne j$)
\be
\frac{\partial \lambda_a}{\partial J_{i j} \partial J_{j i}} =
\frac{m_i m_j}{m_i^2 - m_j^2}
\ee
when $i=a$, $i$ and $j$ are reversed when $j=a$, and it is zero otherwise.
Putting this all together we end up with
\be
\Omega_2 &=&
\frac12\sum_{i,j}^n m_i m_j \frac{\partial^2 \Omega_0}
{\partial m_i \partial m_j}
+\frac14\sum_i^n \left(
m_i \frac{\partial \Omega_0}{\partial m_i}
-m_i^2 \frac{\partial^2 \Omega_0}{\partial m_i^2}
-m_i^2 \Omega_0
\right)
-\frac14 n\nu^2\Omega_0 \cr &&
-\frac12\sum_{i \ne j}^n \frac{m_i m_j}{m_i^2 - m_j^2}
 \left( m_j \frac{\partial \Omega_0}{\partial m_i} 
 - m_i \frac{\partial \Omega_0}{\partial m_j} \right).
\ee
This can be further simplified by using the identity \cite{broweru}
\be
\sum_i^n \left[
(2n-1)m_i \frac{\partial \Omega_0}{\partial m_i}
+m_i^2 \frac{\partial^2 \Omega_0}{\partial m_i^2}
-m_i^2 \Omega_0
\right]
+2 \sum_{i \ne j}^n \frac{m_i m_j}{m_i^2 - m_j^2}
\left( m_j \frac{\partial \Omega_0}{\partial m_i} 
- m_i \frac{\partial \Omega_0}{\partial m_j} \right)
= n\nu^2\Omega_0.\nonumber \\
\ee
As a final result we obtain
\be
\Omega_2 =
-\frac12(n-1)\sum_i^n m_i \frac{\partial \Omega_0}{\partial m_i}
+\frac12 \sum_{i \ne j}^n m_i m_j
 \frac{\partial^2 \Omega_0}{\partial m_i \partial m_j}
-\sum_{i \ne j}^n \frac{m_i m_j}{m_i^2 - m_j^2}
 \left( m_j \frac{\partial \Omega_0}{\partial m_i} 
 - m_i \frac{\partial \Omega_0}{\partial m_j} \right).\nonumber \\
\label{finalo2}
\ee
For $N_f=1$ with sea quark mass $m_1$ and valence quark mass $m_v$ the
expression (\ref{finalo2}) becomes for arbitrary $\nu$
\be
\Omega_2 = -\frac12 \Omega_1
- m_v^2 \Omega_0
-\nu^2 \Omega_0
+ \nu \Omega_1 + 2 \nu \Omega_0
+ 2 m_v I_{1+\nu}(m_v) I_\nu(m_v)
~,
\ee
while for $N_f$ massless flavors and one valence quark  we get
for  $\nu=0 $
\be
\Omega_2 = -\frac12 N_f \Omega_1.
\ee

\end{document}